\documentclass[11pt,a4paper]{article}
\usepackage{jstyle}
\usepackage[bbgreekl]{mathbbol}

\author[a]{Jin-Beom BAE}
\author[b]{\quad Euihun JOUNG}
\author[c]{\quad Shailesh LAL}

\affiliation[a]{Korea Institute for Advanced Study, 85 Hoegiro, Dongdaemun-Gu, Seoul 02455, Korea}
\affiliation[b]{Department of Physics and Research Institute of Basic Science, Kyung Hee University, Seoul 02447, Korea}
\affiliation[c]{LPTHE -- UMR 7589, UPMC Paris 06, Sorbonne Universit{\'e}s,  Paris 75005, France}

\emailAdd{jinbeom@kias.re.kr}
\emailAdd{euihun.joung@khu.ac.kr}
\emailAdd{shailesh@lpthe.jussieu.fr}

\title{\centering One-Loop Free Energy of \\ Tensionless Type IIB String in AdS$_5\times$S$^5$}

\abstract{Considering the zero 't Hooft coupling limit of \mt{\cN=4} super-Yang-Mills theory,
 the exact spectrum of all single-trace operators can be accessed in terms of
the underlying $so(2,4)$ character.
This makes it possible in turn to compute the one-loop free energy
 of the tensionless type IIB string theory in AdS$_5\times$S$^5$ background,
 with help of the recently developed method of \emph{character integral representation of zeta function (CIRZ)}.
 We  calculate first the one-loop free energy of the string states in the \mt{(p-1)}-th Regge trajectory
and find the result to be $p$ times the free energy of a single \mt{\cN=4} Maxwell multiplet. The full one-loop free energy is hence proportional to the divergent series
\mt{\sum_{p=2}^\infty p}\,. The divergence arises as a result of interrupting the regularization procedure in an intermediate stage.
With a reorganization of states, we extract the finite part of free energy
after summing over the Regge trajectories.
This way gives  us a finite result
which is minus of the free energy of the \mt{\cN=4} multiplet.
Hence, this bulk one-loop result matches the $-1$ term in the  \mt{N^2-1}  factor of the boundary result.

 }

\begin{document}

\maketitle

\section{Introduction}

Ever since the proposal of the duality between
four-dimensional \mt{\cN=4} $SU(N)$ super-Yang-Mills   (SYM) 
and Type-IIB superstring theory on 
\mt{AdS_5\times S^5} \cite{Maldacena:1997re,Gubser:1998bc,Witten:1998qj},
there have been innumerous tests on this conjecture. 
Here, we revisit the match of 
free energies of the two theories. 
Considering the boundary geometry as $S^4$ or \mt{S^1\times S^3}\,,
the free energy of the \mt{\cN=4} SYM gives the conformal $a$-anomaly or Casimir energy on $S^3$\,, respectively.
These quantities are protected from renormalization and given exactly by
\ba
	&& F^{\sst SU(N)}_{\sst \cN=4}(S^4)=(N^2-1)\,\log\L\,,
	\nn
	&&F^{\sst SU(N)}_{\sst \cN=4}(S^1_\bbbeta\times S^3)=(N^2-1)\,\frac{3}{16}\,\bbbeta+\cO(\bbbeta^0)\,,
	\label{boundary FE}
\ea
where $\L$ is the UV cut-off and $\bbbeta$ is the radius of the thermal cycle $S^1$\,.
On the other hand, the bulk free energy admits the loop expansion, 
\be
	\frac1G\,\G_{\sst\rm IIB}
	=\frac1G\,S_{\sst\rm IIB}+\G^{\sst (1)}_{\sst\rm IIB}+\cO(G)\,.
\ee 
With the dictionary $1/G=N^2$\,, the leading order $S_{\sst\rm IIB}$ 
reproduces the $N^2$ term of the boundary free energy in the supergravity limit \cite{Henningson:1999xi,Balasubramanian:1999re}. Moving to the next order $\G^{\sst (1)}_{\sst\rm IIB}$\,, one needs to consider one-loop effects in type-IIB string theory in the bulk \cite{Mansfield:2000zw,Mansfield:2003gs,Beccaria:2014xda}. 

The excitations of  type-IIB string in $AdS_5\times S^5$
can be organized into multiplets of
the supergroup $PSU(2,2|4)$ 
\cite{Gunaydin:1998jc,Gunaydin:1998sw,Dolan:2002zh,Bianchi:2003wx,Beisert:2003te,Beisert:2004di}. The 10d massless modes correspond to BPS ones --- the 5d \mt{\cN=8} supergravity multiplet and infinite tower of massive Kaluza-Klein (KK) multiplets \cite{Kim:1985ez} ---
whereas the 10d massive modes correspond to various non-BPS multiplets
whose masses are determined by the string tension. 
The contribution of the massive modes to the one-loop free energy is expected to vanish as the boundary result \eqref{boundary FE} does not depend on the 't Hooft coupling or string tension.
In contrast, the contributions of the 
massless modes do not vanish
(they are studied
first in \cite{Mansfield:2000zw,Mansfield:2003gs}
 and revised in \cite{Beccaria:2014xda}):
\ba
	&&\G^{\sst (1)}_{{\rm\sst KK},p}(AdS_5\times S^5)=p\,\log R\,,
	\nn
	&& \G^{\sst (1)}_{{\rm\sst KK},p}(AdS^\bbbeta_5\times S^5)=p\,\frac{3}{16}\,\bbbeta
	+\cO(\bbbeta^0)\,,
	\label{KK}
\ea
where \mt{p-1} is
the KK level and $R$ is the IR cut-off of the radial coordinate in AdS$_5$\, which maps to the 
boundary UV cut-off $\Lambda$\,.
AdS$^\bbbeta_5$ is the thermal AdS$_5$ with $S^1_\bbbeta\times S^3$ boundary.
Therefore, with the assumption that the contributions of 10d massive states vanish, the full one-loop free energy,
\ba
	&& \G^{\sst (1)}_{\sst\rm IIB}(AdS_5\times S^5)=A\,\log R\,,
	\nn
	&&\G^{\sst (1)}_{\sst\rm IIB}(AdS^\bbbeta_5\times S^5)=A\,\frac{3}{16}\,\bbbeta
	+\cO(\bbbeta^0)\,,
	\label{IIB FE}
\ea
is given with a divergent series $A=\sum_{p=2}^\infty p$\,.
An appropriate regularization scheme 
is needed for a physically meaningful answer. The desirable result is $A =-1$ so that the semi-classical and one-loop results sum up to reproduce exactly the $N^2-1$ factor of the boundary free energy.

\subsection*{Tensionless Limit of Type IIB String Theory}

With the assumption that the massive 10d string modes do not contribute 
to the one-loop free energy, 
we avoid the problem of identifying such multiplets in the string spectrum. In the tensile case, this problem is highly non-trivial as it amounts to calculating 
exact anomalous dimensions for all single-trace operators.
In the tensionless limit, the identification of all these single-trace operators --- hence, the string excitation modes --- 
becomes available in principle.
In this paper, we 
calculate the one-loop free energy of the tensionless type IIB strings in a straightforward and thorough manner,
that is, by properly collecting the contributions of all single-particle states in the string spectrum while making no \textit{a priori} assumptions about the vanishing of the massive mode contribution.
As we shall show, our result reproduces the expression \eqref{IIB FE} with $A=\sum_{p=2}^\infty p$, but now
each $p$ contribution comes from all the states residing in the \mt{(p-1)}-th ``Regge trajectory'',\footnote{The term ``Regge trajectory'' refers to the collection of particles with the same mass versus spin ratio, and 
is useful to organize the string spectrum around flat spacetime.
However, in \emph{tensionless string} of AdS$_5\times$S$^5$\,, the spectrum does not
have this pattern, and we use the term ``Regge trajectory'' for another way to organize 
the entire string spectrum. The precise definition will be given after the equation \eqref{tr phi}.} which contains in fact the \mt{(p-1)}-th KK states.

Our work is motivated by the recent success of the dualities \cite{Sezgin:2002rt,Klebanov:2002ja} between higher spin gravity \cite{Vasiliev:1990en} and vector model conformal field theories,
where the single-particle/trace states are simple enough to allow a direct evaluation of bulk one-loop free energy
\cite{Giombi:2013fka,Giombi:2014iua,Giombi:2014yra} (see \cite{Beccaria:2014zma,Basile:2014wua,Beccaria:2015vaa,Beccaria:2016tqy,Pang:2016ofv,Gunaydin:2016amv,Giombi:2016pvg,Brust:2016xif,Bae:2016xmv} for more recent works).
To extend the success of vector models to adjoint models,
the authors of the current article have worked out two string-like bulk theories, one dual to free
scalar \cite{Bae:2016rgm} and the other dual to free Yang-Mills \cite{Bae:2016hfy} with $SU(N)$ adjoint symmetry.
The key ingredient in our previous work is the use of Character Integral Representation of Zeta function (CIRZ)
devised in \cite{Bae:2016rgm}, which allows us to calculate the free energy with the information encoded in the character only.
Also in the current paper, the CIRZ method enables us to perform the computation in a straightforward manner.

\section{Single-Particle States/Single-Trace Operators}
\label{sec:ST}

In the tensionless limit ($\a'\to \infty$ while $L_{\rm\sst AdS}$ finite) of type IIB string theory in the bulk
\cite{HaggiMani:2000ru,Sundborg:2000wp,wittenHS,Tseytlin:2002gz,Karch:2002vn,Gopakumar:2003ns,Bonelli:2003zu,Gopakumar:2004qb,Gopakumar:2004ys,Gopakumar:2005fx,Aharony:2006th,Yaakov:2006ce,Aharony:2007fs,TseytlinHS3,Amado:2016pgy}, the 't Hooft coupling of the boundary $\cN=4$ SYM goes to
zero,
\be
	\l=N\,g_{\rm\sst YM}^2=\left(\frac{L_{\rm\sst AdS}^2}{\a'}\right)^2 \to 0\,,
\ee
hence all the operators become free from any anomalous dimensions.
In this case, the spectrum of single-trace operators can be
exactly identified by group theoretical and combinatorial methods \cite{Polyakov:2001af,Spradlin:2004pp,Barabanschikov:2005ri,Newton:2008au}. 
Via the AdS/CFT correspondence, 
we identify this with the spectrum of single-particle states in the bulk
and consider the corresponding bulk theory,  following \cite{Sundborg:1999ue}, as our \emph{working definition} for tensionless string theory.
By exploiting this definition, one can collect and sum over the one-loop free energies of all single-particle states in the string spectrum \cite{Bae:2016hfy,Bae:2016rgm}. 

For the group theoretical identification of the operators
made by $p$ insertions of field in a single trace,
\be
	\tr(\Phi_1\cdots \Phi_p)\,,
	\qquad \Phi_i\,:\, \textrm{any field (or its derivative) of }\cN=4\textrm{ SYM}\,,
	\label{tr phi}
\ee 
we need to consider the $p$-th tensor product of the \mt{\cN=4} Maxwell multiplet and project it to the singlet under cyclic permutations.
Each of the conformal primaries lying in this tensor product space is identified via the AdS/CFT correspondence to a field in AdS$_5$ as mentioned above. 
Henceforth, for notational convenience, we define the $(p-1)$-th Regge trajectory to be 
the set of all such fields.

The issue of tensor products and decompositions can be conveniently analyzed in terms of the $so(2,4)$ character.
The cyclicity is required by the property of trace operation,
and handling the cyclic projection is equivalent to the combinatorial problem
of counting the number of
different necklaces with $p$ beads where each bead corresponds to
a state --- either primary or descendant --- of \mt{\cN=4} Maxwell multiplet.
Polya's enumeration theorem solves this problem
by making use of the cyclic index.

\subsection*{Partition function of $\cN=4$ Maxwell multiplet}

The cyclic index can be used to account the tensor products
of $so(2,4)$ representation $\cH$,
if the index variables are replaced by $\tr_{\cH}(g^k)$
with $g\in SO(2,4)$\,.
In this way, the cyclic index of a bosonic system gives nothing but the $so(2,4)$ character.
For more general systems with fermionic degrees of freedom, we should consider the weighted partition function \cite{Beisert:2003te,Bianchi:2006ti},
\be
	\cZ_\cH(q,x_1,x_2)=\tr_\cH\left((-1)^F\,q^{M_{05}}\,x_1^{M_{12}}\,x_2^{M_{34}}\right),
	\label{index}
\ee
where $M_{ab}$ are the generators of $so(2,4)$ and $F$ is the fermion number.
Even though we refer to this as partition function
following previous references, 
it should be distinguished from the exponential of the one-loop free energy that we shall calculate in the next section.
It can be equally viewed as a generalized Witten index.
For an irreducible representation of $so(2,4)$\,, it is 
proportional to the character up to a sign factor:
\be
	\cZ_{\cD(\D,[j_+,j_-])}=(-1)^{2(j_++j_-)}\,\chi_{\cD(\D,[j_+,j_-])}\,,
\ee
and for the \mt{\cN=4} Maxwell multiplet, it is given by
\be
	\cZ^{\phantom{g}}_{\sst \cN=4}=
	\chi_{\sst 1}
	-4\,\chi_{\sst \frac12}
	+6\,\chi_{\sst 0}\,.
\ee
Here, $\chi_s$ are the characters of massless spin $s$ representations given in \cite{Dolan:2005wy}
\ba
	\chi_{\sst 1}(\b,\a_1,\a_2)\eq \chi_{\cD\left(2,[1,0]\right)}(\b,\a_1,\a_2)+\chi_{\cD\left(2,[0,1]\right)}(\b,\a_1,\a_2) \nn
	\eq \frac{e^{-\beta }\,(\cosh\b -\cos \a_1-\cos\a_2)+
	\cos \a_1\, \cos\a_2}{(\cos \a_1-\cosh \beta ) (\cos \a_2-\cosh \beta )}\,,
\ea
\ba
\chi_{\sst \frac12}(\b,\a_1,\a_2)\eq 
\chi_{\cD\left(\frac32,\left[\frac12,0\right]\right)}(\b,\a_1,\a_2)
	+\chi_{\cD\left(\frac32,\left[0,\frac12\right]\right)}(\b,\a_1,\a_2)\nn
\eq\frac{2\, \cos\frac{\a_1}{2}\,\cos\frac{\a_2}{2} \,\sinh\frac{\beta }{2}}{(\cos \a_1-\cosh \beta ) (\cos \a_2-\cosh \beta )}\,,
\ea
\be
	\chi_{\sst 0}(\b,\a_1,\a_2)=\chi_{\cD(1,[0,0])}(\b,\a_1,\a_2)=
	\frac{\sinh\beta}{2 (\cos\a_1-\cosh\beta) (\cos\a_2-\cosh\beta)}\,.
\ee
We thus obtain
\ba
&&\cZ^{\phantom{g}}_{\sst \cN=4}(\b,\a_1,\a_2)\\
&&=
\frac{e^{-\b}(\cosh\b-\cos\a_1-\cos\a_2)
+\cos \a_1 \cos \a_2
-8\,\cos\frac{\a_1}{2}\, \cos\frac{\a_2}{2}\, \sinh\frac{\beta }{2}
+3\,\sinh\beta}
{(\cos \a_1-\cosh\beta) (\cos\a_2-\cosh\beta)}\,,\nonumber
\ea
where the variables $\b,\a_1,\a_2$ are related to the ones in \eqref{index} by $q=e^{-\b}$\,, $x_1=e^{i\,\a_1}$
and $x_2=e^{i\,\a_2}$\,.

\subsection*{Partition function of the $(p-1)$-th Regge trajectory of 
tensionless type IIB string}

As mentioned before, the partition function \eqref{index} takes 
the fermionic statistics properly into account
in making cyclic projection \cite{Beisert:2003te,Bianchi:2006ti}.
This quantity is computed over
the cyclic tensor product of \mt{\cN=4} Maxwell multiplet is given by
\ba
		&& \cZ^{\sst {\rm cyc}(p)}_{\sst \cN=4}(q,x_1,x_2)
		= \frac1{p}\sum_{k|p}
	\varphi(k)\left(\cZ_{\sst \cN=4}(q^{k},x_1^k,x_2^k)\right)^{\frac pk}\nn
	&&= \cZ_{\sst {\rm IIB}(p)}(q,x_1,x_2)=
	\sum_{\D,j_+,j_-}
	N^{\sst{\rm IIB}(p)}_{\cD(\D,[j_+,j_-])}\,
	\cZ_{\cD(\D,[j_+,j_-])}(q,x_1,x_2)\,.
	\label{IIB p index}
\ea
Here, $\varphi(k)$ is the Euler totient function, which counts 
the number of relative primes of $k$ in $\{1,2,\ldots, k\}$\,,
and 
$N^{\sst{\rm IIB}(p)}_{\cD(\D,[j_+,j_-])}$ is the number
of the $\cD(\D,[j_+,j_-])$ representation in the $(p-1)$-th Regge trajectory
of the type IIB string theory in the tensionless limit.

\section{One-Loop Free Energy of Tensionless Type IIB String}
\label{sec:FE}

The one-loop free energy of type IIB string in $AdS_5\times S^5$ background is  simply 
the sum of the individual free energies
for all string states.
Therefore, as a matter of principle, 
it is possible to obtain the full quantity in the tensionless limit since the 
information about the exact multiplicities 
$N^{\sst{\rm IIB}(p)}_{\cD(\D,[j_+,j_-])}$ 
is encoded in the partition function \eqref{IIB p index}.
However, in practice, it is \emph{not} possible to extract analytic 
expressions for all these multiplicities,
needless to mention about the eventual resummation.

\subsection{Character Integral Representation for Zeta function}
 
To overcome this problem, a new method of computing one-loop free energy has been devised in \cite{Bae:2016rgm}.
This method, referred to as Character Integral Representation of Zeta function (CIRZ),
makes it possible to obtain the spectral zeta function associated to the one-loop free energy as a certain integral of 
the $so(2,4)$ character.
In this way, we can handle the contributions of all single-particle states 
without identifying their actual content.
We refer \cite{Bae:2016rgm} for the detailed derivation of this method.
In our previous works \cite{Bae:2016rgm,Bae:2016hfy,Bae:2016xmv}, we have considered only bosonic systems, hence the method involved
the $so(2,4)$ character.
In order to generalize the method to incorporate fermionic states,
it is sufficient to replace the $so(2,4)$ character by the partition function \eqref{index}
since the one-loop free energies of bosons and fermions in odd-dimensional AdS are given in a  completely analogous 
form but only with an overall  minus sign factor.
In below, we summarize the result of the CIRZ method.

\subsection*{AdS$_5$ with S$^4$ boundary}

The one-loop free energy of a spectrum $\cH$ in $AdS_5$ with $S^4$ boundary is 
\be
	\G^{\sst (1)\,{\rm ren}}_{\cH}(AdS_5)=
	\log R\left(\gamma_{\mathcal{H}|2}+\gamma_{\mathcal{H}|1}+\gamma_{\mathcal{H}|0}\right).
	\label{FE}
\ee
Here, $R$ is  the IR cut-off for the radial coordinate in $AdS_5$ \cite{Henningson:1998gx} 
and $\gamma_{\mathcal{H}|n}$ are given by
\be
	\int_0^\infty d\b\,\frac{\left(\frac{\b}2\right)^{2(z-1-n)}}{\G(z-n)}\,
	f_{\cH|n}(\b)
	=-2\,\gamma_{\cH|n}+\mathcal{O}(z)\,,
	\label{FE real int}
\ee
where $f_{\cH|n}(\b)$ are determined by the partition functions $\mathcal{Z}_{\cH}$ as 
\begin{equation}\label{f H}
\begin{split}
f_{\mathcal{H}|2}(\b)&= {\sinh^4{\tfrac\beta2}\over 2}\,\cZ_{\mathcal{H}}\!\left(\beta,0,0\right),\\
f_{\mathcal{H}|1}(\b)&= \sinh^2{\tfrac\beta2}\left[{\sinh^2{\tfrac\beta2}\over 3}-1-\sinh^2{\tfrac\beta2}\left(\partial_{\alpha_1}^2 +\partial_{\alpha_2}^2\right)\right]\cZ_{\mathcal{H}}\!\left(\beta,\alpha_{1},\alpha_{2}\right)\bigg|_{\alpha_{i}=0},\\
f_{\mathcal{H}|0}(\b)&=\left[1 + {\sinh^2\tfrac\beta2\left(3-\sinh^2\tfrac\beta2\right)\over 3} \left(\partial_{\alpha_1}^2 +\partial_{\alpha_2}^2\right)\right.\\&\qquad\left. -{\sinh^4\tfrac\beta2\over 3}\left(\partial_{\alpha_1}^4-12\,\partial_{\alpha_1}^2\partial_{\alpha_2}^2+\partial_{\alpha_2}^4\right)\right]\cZ_{\mathcal{H}}\!\left(\beta,\alpha_{1},\alpha_{2}\right)\bigg|_{\alpha_{i}=0}.
\end{split}
\end{equation}
The variables $\b,\a_1,\a_2$ are related to the ones in \eqref{index} by $q=e^{-\b}$\,, $x_1=e^{i\,\a_1}$
and $x_2=e^{i\,\a_2}$\,.
When the functions $f_{\cH|n}$ do not have any singularity apart from poles at origin,
we can deform the integral \eqref{FE real int} to a contour one and obtain
\begin{equation}\label{gammahn}
\gamma_{\mathcal{H}|n} = -\left(-4\right)^n n! \oint {d\beta\over 2\pi i}\,{f_{\mathcal{H}|n}(\b)\over\beta^{2\left(n+1\right)}}\,,
\end{equation}
where the contour encircles the origin in the anticlockwise direction.

\subsection*{Thermal AdS$_5$ with S$^1\times$S$^3$ boundary}

When the background is the thermal $AdS_5^\bbbeta$ with $S^1_\bbbeta\times S^3$ boundary
($\bbbeta$ is the radius of $S^1$)
the one-loop free energy reads
\be
	\G^{\sst (1)\,\rm ren}_{\cH}(AdS_5^\bbbeta)
	=\bbbeta\,\cE_{\cH}+\cO(\bbbeta^0)\,,
	\label{th FE}
\ee
where  $\cE_{\cH}$ is  the Casimir energy given by 
\be
	\int_0^\infty \frac{d\b\,\b^{z-1}}{\Gamma(z)}\,\cZ_\cH(\b,0,0)
	=2\,\cE_{\cH}+\cO(z+1)\,.
	\label{E real}
\ee
If $\cZ_\cH(\b,0,0)$ does not have any singularity apart from poles at the origin,
we can deform the integral such that we obtain the Casimir energy as
\be
	\cE_{\cH}
	=-\frac12\,\oint \frac{d\b}{2\,\pi\, i\,\b^2}\,\cZ_\cH(\b,0,0)\,,
	\label{E H}
\ee
where the contour is again an anticlockwise circle around the origin. 

\bigskip

We now compute the  one-loop free energies
\eqref{FE} and \eqref{th FE} for the tensionless string.
To begin with, we organize  the string states  according to the level of Regge trajectories 
and compute the one-loop free energy at each trajectory.

\subsection{One-Loop Free Energy of the $(p-1)$-th Regge Trajectory}

The partition function of the $(p-1)$-th Regge trajectory is
\be
 \cZ_{\sst {\rm IIB}(p)}(\b,\a_1,\a_2)=
\frac1{p}\sum_{k|p}
	\varphi(k)\left[\cZ_{\sst \cN=4}(k\,\b,k\,\a_1,k\,\a_2)\right]^{\frac pk}\,,
	\label{p index}
\ee
where the partition function of \mt{\cN=4} Maxwell multiplet has the form of
\ba
&&\cZ_{\sst \cN=4}(\b,\a_1,\a_2)=\\
&&=\,\frac{e^{-\b}(\cosh\b-\cos\a_1-\cos\a_2)
+\cos \a_1 \cos \a_2
-8\,\cos\frac{\a_1}{2}\, \cos\frac{\a_2}{2}\, \sinh\frac{\beta }{2}
+3\,\sinh\beta}
{(\cos \a_1-\cosh\beta) (\cos\a_2-\cosh\beta)}\,.
\nonumber
\ea
We first note that  both one-loop free energies
\eqref{FE} and \eqref{th FE} of the \mt{(p-1)}-th Regge trajectory
are given 
by a few series coefficients of $\cZ_{\sst{\rm IIB}(p)}(\b,0,0)$
or $f_{\sst{\rm IIB}(p)|n}(\b)$
 because the latter do not have any singularity near origin
 hence the residue theorem can be applied to the contour integrals \eqref{gammahn} and \eqref{E H}.
Since the latter is a bit more involved, let us begin with the former case,
that is, the Casimir energy.
We first consider the series expansion of the partition function $\cZ_{\sst \cN=4}(\b,0,0)$\,:
\be
	\cZ_{\sst \cN=4}(\b,0,0)=
	\frac{2+6\,e^{\frac{\b}2}}{(1+e^{\frac\b2})^3}
	=1+a\,\b+\cO(\b^3)\,,
	\qquad a=-\frac38\,.
\ee
It is worth to remark the speciality of the above expression:
the partition function of a generic multiplet consisting of $n_1$ spin 1,
$n_\frac12$ spin 1/2 and $n_0$ spin 0 has the series expansion,
\ba
	&& \cZ_{\{n_1,n_{\frac12},n_0\}}(\b,0,0)=\nn
	&&
	=\,\frac{2\,n_0-4\,n_\frac12+4\,n_1}{\beta ^3}+\frac{n_\frac12-4\,n_1}{2\, \beta }+n_1-\frac{4\,n_0+17\,n_\frac12+88\,n_1}{480}\,\beta  +\cO(\b^3)\,,
	\label{Z gen}
\ea
hence involves the $\b^{-3}$ and $\b^{-1}$ terms. 
Note that the $\b^{-3}$ term vanishes in any supersymmetric theory,
but its physical meaning is not clear to the authors.
When the SUSY is the maximal $\cN=4$ one, 
not only the $\b^{-3}$ term but also the $\b^{-1}$ term (hence all negative powers)
vanish.
In this case, we find  a drastic simplification in the expansion formula for $\left[ \cZ_{\sst \cN=4}(k\,\b,0,0)\right]^{\frac{p}{k}}$\,:
\be
	\left[ \cZ_{\sst \cN=4}(k\,\b,0,0)\right]^{\frac{p}{k}}=
	1-\,\frac38\,p \, \b+\cO(\b^2)\,,
	\label{pk exp}
\ee
which appear in the partition function of the cyclic tensor product \eqref{p index}.
Moreover, since \eqref{pk exp} is independent of $k$\,,
we can perform the summation over $k$ in \eqref{p index} by using the identity,
\be
	\frac1p\,\sum_{k|p}\,\varphi(k)=1\,.
	\label{ET id}
\ee
Finally, we find that the series expansion of $\cZ_{\sst {\rm IIB}(p)}(\b,0,0)$ itself is given by 
the right hand side of \eqref{pk exp}.
Hence,  the thermal AdS$_5$ one-loop free energy is 
\be
	\Gamma^{\sst(1)\,\rm ren}_{\sst{\rm IIB}(p)}(AdS^\bbbeta_5\times S^5)
	=-\frac12\,p\,a\,\bbbeta+\cO(\bbbeta^0)=
	\frac{3\,p}{16}\,\bbbeta+\cO(\bbbeta^0)\,,
	\label{th FE p}
\ee
for the tensionless string states in the $(p-1)$-th Regge trajectory. 
This result coincides with that of the $(p-1)$-th KK level \eqref{KK}.

Let us now move on to the one-loop free energy in $AdS_5$ with $S^4$ boundary.
In this case, the partition function $\cZ_{\sst{\rm IIB}(p)}(\b,\a_1,\a_2)$ enters in the computation through 
the functions $f_{\sst{\rm IIB}(p)|n}(\b)$ defined in \eqref{f H}.
From the residue theorem, the $\gamma_{\sst{\rm IIB}(p)|n}$ coefficient  \eqref{gammahn}
of the one-loop free energy \eqref{FE} depends only on 
the $\b^{2n+1}$ coefficient of $f_{\sst{\rm IIB}(p)|n}(\b)$\,.
However, since the latter combines $\cZ_{\sst{\rm IIB}(p)}(\b,\a_1,\a_2)$ with
hyperbolic functions, one may expect that the one-loop free energy depends on 
higher series coefficients of the partition function $\cZ_{\sst{\rm IIB}(p)}(\b,\a_1,\a_2)$.
However, as we shall show now, this kind of complications do not happen due to the speciality 
of the \mt{\cN=4} partition function.
About the series expansion in $\a_1$ and $\a_2$, only the terms up to $\cO(\a^6)$ are relevant,
hence for the building block  $\cZ_{\sst \cN=4}(\b,\a_1,\a_2)$\,, it is  sufficient to consider 
the series expansion to the same order:
\ba
\cZ_{\sst \cN=4}(\b,\a_1,\a_2)\eq \frac{2+6\,e^{\frac{\b}2}}{(1+e^{\frac\b2})^3}
+\frac{\sinh^4\frac{\b}4}{\sinh^5\frac{\b}2}\left(\a_1^2+\a_2^2\right)
-\frac{3+\cosh\b}{2048\,\sinh^3\frac{\b}4\,\cosh^7\frac{\b}4}\,\a_1^2\,\a_2^2\nn
&&\quad +\,\frac{5-\cosh\frac\b2}{3072\,\sinh\frac{\b}4\,\cosh^7\frac{\b}4}\left(\a_1^4+\a_2^4\right)
+\cO(\a^6)\,.
\ea
We can expand the above in $\b$ to get
\ba
\cZ_{\sst \cN=4}(\b,\a_1,\a_2)\eq
1+a\,\b+\cO(\b^3)
+\left(b\,\b+\cO(\b^3)\right)\frac{\a_1^2+\a_2^2}{\b^2}\nn
&&
+\left(c\,\b+\cO(\b^3)\right)\frac{\a_1^2\,\a_2^2}{\b^4}
+\cO(\b^3)\,\frac{\a_1^4+\a_2^4}{\b^4}+\cO(\a^6)\,,
\label{N=4 exp a b}
\ea
where $a, b, c$ are constants given by
\be
	a=-\frac38\,,\qquad
	b=\frac18\,,\qquad c=-\frac18\,.
\ee
Again the absence of negative $\b$ powers  in $\cZ_{\sst \cN=4}(\b,0,0)$
allows us to obtain a compact series expansion formula for 
$\left[\cZ_{\sst \cN=4}(k\,\b,k\,\a_1,k\,\a_2)\right]^\frac{p}{k}$\,:
\ba
&&\left[\cZ_{\sst \cN=4}(k\,\b,k\,\a_1,k\,\a_2)\right]^\frac{p}{k}=
1+p\,a\,\b+\cO(\b^2)
+\left(p\,b\,\b+\cO(\b^2)\right)\frac{\a_1^2+\a_2^2}{\b^2}\nn
&&\hspace{60pt}
+\left(p\,c\,\b+\cO(\b^2)\right)\frac{\a_1^2\,\a_2^2}{\b^4}
+\cO(\b^2)\,\frac{\a_1^4+\a_2^4}{\b^4}+\cO(\a^6)\,.
\label{I exp}
\ea
Moreover, the $k$ dependence disappears again in the above expression
and the summation over $k$ in $\cZ_{\sst {\rm IIB}(p)}(\b,\a_1,\a_2)$ \eqref{p index}
can be performed with the identity \eqref{ET id}.
As a result, the partition function $\cZ_{\sst {\rm IIB}(p)}(\b,\a_1,\a_2)$ has exactly
the same series expansion as the right hand side of \eqref{I exp}.
Even though $f_{\sst{\rm IIB}(p)|n}(\b)$ involves a few hyperbolic functions,
eventually the coefficients $\g_{\sst{\rm IIB}(p)|n}$
are all determined from the series coefficients of $\cZ_{\sst {\rm IIB}(p)}$ that we have just identified.
In the end, we obtain
\be
	\g_{\sst{\rm IIB}(p)|2}=-p\,a\,,
	\qquad
	\g_{\sst{\rm IIB}(p)|1}=-p\,(a+b)\,,
	\qquad
	\g_{\sst{\rm IIB}(p)|0}=-p\,(a+b+c)\,,
\ee
and combining these, the AdS$_5$ one-loop free energy is obtained as
\be
	\Gamma^{\sst(1)\,\rm ren}_{\sst{\rm IIB}(p)}(AdS_5\times S^5)
	=-p\,(3\,a+2\,b+c)\,\log R=p\,\log R\,,
	\label{FE p}
\ee
for the 
tensionless string states in the \mt{(p-1)}-th Regge trajectory.
Again, this result coincides with that of the \mt{(p-1)}-th KK level \eqref{KK}.

\subsection{Full One-Loop Free Energy}
\label{sec: FFE}

In the previous section, we have calculated the 
one-loop free energies $\Gamma^{\sst(1)\,\rm ren}_{\sst{\rm IIB}(p)}$ of the tensionless type IIB string 
in a given  Regge trajectory.
The full one-loop free energy is  their sum over trajectories,
\be
	\Gamma^{\sst(1)\,\rm ren}_{\sst{\rm IIB}}=
	\sum_{p=2}^\infty\Gamma^{\sst(1)\,\rm ren}_{\sst{\rm IIB}(p)}\,.
\ee
With the results \eqref{FE p} and \eqref{th FE p}, 
the above gives \eqref{IIB FE} with the divergent series $A=\sum_{p=2}^\infty p$.
In order to reproduce the physically desirable result, there should exist an appropriate regularization scheme
giving rise to 
\be
	\sum_{p=1}^\infty p=0\,.
	\label{p sum}
\ee
In \cite{Beccaria:2014xda}, it has been shown that 
the standard zeta function regularization
is compatible with the insertion of $e^{-\e\,M_{05}}$ 
inside the trace.
With the latter regularization, if we take the contribution of the entire KK towers of 10d supergravity keeping 
the regularization parameter $\e$ finite and 
send $\e$ to zero in the final stage, then we recover
the result which is consistent with the scheme \eqref{p sum}.
Here, the key idea is that the extraction of the finite part 
from the regulator dependent quantity
should be done after the summation over KK levels.
Coming back to the standard zeta function regularization 
which the current paper is relying on,
we can also attempt to extract the finite part of the result
 after performing the summation over the entire Regge 
trajectories. 
In our method of CIRZ, this would correspond to 
consider the full partition function,
\be
	\cZ_{\sst\rm IIB}
	=\sum_{p=2}^\infty \cZ_{\sst{\rm IIB}(p)}\,,
\ee
before performing various integrals.
In fact, by using the  re-summation,
\be
	\sum_{p=1}^{\infty}\,\sum_{k|p} \Big(\cdots\Big)=\sum_{k=1}^{\infty}\,\sum_{n=1}^{\infty} \Big(\cdots\Big)
	\qquad [p=k\,n]\,,
	\label{reorg}
\ee
we can consider the partition function $\cZ_{\sst\rm IIB}$ as
a different series \cite{Sundborg:1999ue}, 
\be
	\cZ_{\sst\rm IIB}(\b,\a_1,\a_2)
	=-\cZ_{\sst \cN=4}(\b,\a_1,\a_2)
-\sum_{k=1}^{\infty}\,\frac{\varphi(k)}{k}\,\log\!\left[1-\cZ_{\sst \cN=4}(k\,\b,k\,\a_1,k\,\a_2)\right],
\label{cyc}
\ee
as a result of the summation over $n$,
which can be viewed as 
an effective sum over the Regge trajectories.

Now let us apply the CIRZ method to the
formula \eqref{cyc}.
We know already the contribution of $-\cZ_{\sst \cN=4}$ to $\Gamma^{\sst(1)\,\rm ren}_{\sst{\rm IIB}}$
 hence move to the series part.
Dealing with the summation over $k$ is technically prohibitive,
but since the summand with fixed $k$ 
contains the contributions
from the entire Regge trajectories, 
this alternative organization of spectrum \eqref{cyc}
may have a better chance to provide 
a convergent series for one-loop free energy.\footnote{
Note however the reorganization \eqref{reorg} does not admit a precise physical interpretation.
In fact, the integrand with fixed $k$ cannot be considered as the partition function over a certain vector space.
Hence, the role of symmetries (conformal symmetry or supersymmetry) on a fixed $k$ is not clear.
}
With this idea in mind, we proceed to the computation 
for each summand with help of CIRZ method.
Again,
all the relevant information is again encoded in the series expansion of the summand,
\ba
	&&\log\!\left[1-\cZ_{\sst \cN=4}(k\,\b,k\,\a_1,k\,\a_2)\right]
=\log\frac{3}8+\log(k\,\b)
+\cO(\b^2)
+\left(-\frac13+\cO(\b^2)\right) \frac{\a_1^2+\a_2^2}{\b^2}\nn
&&\hspace{70pt}
+\left(\frac{2}{9}+\cO(\b^2)\right)\frac{\a_1^2\,\a_2^2}{\b^4}
+\left(-\frac1{18}+\cO(\b^2)\right)\frac{\a_1^4+\a_2^4}{\b^4}
+\cO(\a^4)\,.
\label{log exp}
\ea
Following the contour integral prescription and taking the $\b^{2n+1}$ coefficient from 
the function $f_{\sst \cH|n}$\,,
we can check that the above,
that is the summand in \eqref{cyc} with fixed $k$, does not give any contribution to the one-loop free energy.
Therefore in this prescription,
the full one-loop free energy is given by minus times
that of \mt{\cN=4} Maxwell multiplet:
\begin{equation}
    \G^{\sst (1)\,\rm ren}_{\rm\sst IIB}=-\G^{\sst (1)\,\rm ren}_{\sst \cN=4}= -F_{\sst \cN=4}^{\sst U(1)}\,,
    \label{the result}
\end{equation}
which reproduces perfectly the $-1$ of the factor $N^2-1$ in \eqref{boundary FE}.

\subsubsection*{Branch cut contribution}

One may wonder why we could simply neglect the $\log(k\,\b)$ term even though its branch cut prevents the
use of the contour integral description in the CIRZ method.
In the following, we shall prove
that the $\log(k\,\b)$ term is actually irrelevant.
To show that, we first subtract 
\be
	\cZ_{\sst {\rm mod},k}(\b)=\log[\tanh(k\,\b)]\,,
	\label{I mod}
\ee
from $\log\!\left[1-\cZ_{\sst \cN=4}(k\,\b,k\,\a_1,k\,\a_2)\right]$
in order to remove the  $\log(k\,\b)$ term in the expansion \eqref{log exp}.
To compensate what is subtracted, we have to put back the contribution of
$\cZ_{\sst {\rm mod},k}(\b)$.
Since the latter is a relatively simple function, one can explicitly check that it does not contribute to the full one-loop free energy. 

Let us consider first the one-loop free energy in thermal AdS$_5$ background with $S^1_\bbbeta\times S^3$ boundary.
Due to the branch cut of the logarithm in
\eqref{I mod}, we need to consider the real line integral representation \eqref{E real},
\be
	\tilde\chi_{\sst{\rm mod},k}(z)=
	\int_0^\infty \frac{d\b\,\b^{z-1}}{\Gamma(z)}\,
	\log\!\left[\tanh(k\,\b)\right],
\ee
instead of the contour one.
The one-loop free energy gets the contribution of $\frac12\,\tilde\chi_{\sst {\rm mod},k}(-1)\,\bbbeta$ but since
\be
	\tilde\chi_{\sst{\rm mod},k}(z)=
	-(4\,k)^{-z}\,\zeta\!\left(z+1,\frac12\right)
	=\cO\!\left(z+1\right),
\ee
this does not affect the result
of the one-loop free energy.

Moving now to the case of the AdS$_5$ background with $S^4$ boundary,
we first find that $\cZ_{\sst {\rm mod},k}$ does not depend on $\a_1, \a_2$\,,
hence the functions $f_{\sst {\rm mod},k|n}$ have simple form,
\ba
	f_{\sst{\rm mod},k|2}(\b)\eq \frac{\sinh^4\frac{\b}2}2\,\log[\tanh(k\,\b)]\,,\nn
	f_{\sst{\rm mod},k|1}(\b)\eq \sinh^2\frac{\b}2\left(\frac{\sinh^2\frac{\b}2}3-1\right)\log[\tanh(k\,\b)]\,,\nn
	f_{\sst{\rm mod},k|0}(\b)\eq \log[\tanh(k\,\b)]\,.
\ea
In order to extract the coefficients $\g_{\sst{\rm mod},k|n}$ from the above,
we need to consider the real line integral \eqref{FE real int} 
because the branch cut of the above functions prevent us to use the contour integral \eqref{gammahn}.
They can be recast in terms of 
\be
    I_k(z,a)=
    	\int_0^\infty d\b\,\frac{\left(\frac{\b}2\right)^{2(z-1)}}{\G(z)}\,
    	e^{-a\,\b}\,\log[\tanh(k\,\b)]\,,
\ee 
as follows:
\ba
	&&\int_0^\infty d\b\,\tfrac{\left(\frac{\b}2\right)^{2(z-1)}}{\G(z)}\,
	f{\sst{\rm mod},k|0}(\b)
	= I_k(z,0)\,,
	\label{ff1} \\
	&& \int_0^\infty d\b\,\tfrac{\left(\frac{\b}2\right)^{2(z-2)}}{\G(z-1)}\,f_{\sst{\rm mod},k|1}(\b)=\nn
	&&=\,\frac58\,I_k(z-1,0)-\frac{I_k(z-1,1)+I_k(z-1,-1)}3
	+\frac{I_k(z-1,2)+I_k(z-1,-2)}{48}\,,
	\label{ff2}\\
	&&\int_0^\infty d\b\,\tfrac{\left(\frac{\b}2\right)^{2(z-3)}}{\G(z-2)}\,f_{\sst{\rm mod},k|2}(\b)=\nn
	&&=\,\frac3{16}\,I_k(z-2,0)-\frac{I_k(z-2,1)+I_k(z-2,-1)}8
	+\frac{I_k(z-2,2)+I_k(z-2,-2)}{32}\,.\quad
	\label{ff3}
\ea
The integral $I_k(z,a)$ is divergent for \mt{a<-2k} in the large $\b$ region. This can be interpreted as IR divergence
and can be regularized by analytic continuation in $a$\,.
We can evaluate $I_k(z,a)$ by series expanding the $\log[\tanh(k\,\b)]$
term in $e^{-2k\,\b}$ as
\be 
    I_k(z,a)
    =-\frac{4^{2-3z}\,k^{1-2z}}{\G(z)}\sum_{m=0}^\infty
    \frac{\G(2z+m-1)\,\zeta(2z+m,\frac12)}{m!}\left(-\frac{a}{4\,k}\right)^m\,.
\ee
Series expanding the above around $z=0,-1,-2$, we find
\ba
    I_k(z,a)\eq\left(\frac1z+\g-2\,\log(2\,k)\right)a+\cO(z)\,, \\
    I_k(z-1,a)\eq -\frac{8\,k^2}3\,a
     -\frac23\left(\frac1z+\g-2\,\log(2\,k)-1\right)a^3+\cO(z)\,,\\
    I_k(z-2,a)\eq \frac{224\,k^4}{45}\,a
    +\frac{32\,k^2}9\,a^3
    +\frac4{15}\left(\frac1z+\g-2\,\log(2\,k)-\frac32\right)a^5+\cO(z)\,,
\ea
where $\g$ is the Euler-Mascheroni constant.
Since the above are all odd functions in $a$, the integrals \eqref{ff1}, \eqref{ff2}
and \eqref{ff3} all vanish up to $\cO(z)$ terms.
Therefore, we find that the modification part $\cZ_{\sst{\rm mod},k}(\b)$ \eqref{I mod} does not give any contribution to the one-loop free energy in $AdS_5$ with $S^4$ boundary and the result \eqref{the result} still stands.

\section{Conclusion}
\label{sec:discussion}

In this paper, we have calculated the one-loop free energy
of tensionless type-IIB string theory in $AdS_5\times S^5$ background
by making use of the CIRZ method and the partition function of the boundary 
\mt{\cN=4} theory whose 't Hooft coupling vanishes.
For a fixed but arbitrary Regge trajectory, we could analytically calculate 
the one-loop free energy thanks to the special form of the partition function
of the \mt{\cN=4} Maxwell multiplet. 
At the first place, we showed that the result is proportional to the trajectory number, hence
the full quantity leads to a linearly divergent series.
For a proper treatment of this divergence, we changed the organization of string states
such that we can first perform the effective trajectory sum
before sending the regularization parameter to zero.
In this way, we obtained a finite result which is consistent with the prediction of the holographic conjecture.

We finally note that the simplifications observed in this paper for free energy computations for the $AdS_5$ superstring are somewhat in contrast to the generic situation in higher-spin theory where supersymmetry is not often useful. Indeed even in resolving the tensionless limit of string theory, supersymmetry has been of limited use. 
Nonetheless, the computations here are analytically possible only when the CFT spectrum
coincides with the  \mt{\cN=4} Maxwell multiplet,
thanks to the cancellation of the negative $\b$ powers in \eqref{Z gen}. 
It would be interesting to understand this occurrence better as it might shed new light on the interplay between higher-spin symmetry and supersymmetry.

\acknowledgments

We would like to thank Rajesh Gopakumar, R Loganayagam 
and Arkady Tseytlin for helpful discussions. The work of EJ was supported in part by the National Research Foundation of Korea through the grant NRF2014R1A6A3A04056670 and the Russian Science Foundation grant 14-42-00047 associated with Lebedev Institute. 
The work of SL is supported by the Marie Sklodowska Curie Individual Fellowship 2014. SL would like to thank ICTS-TIFR and the Department of Physics and Astrophysics, University of Delhi for hospitality while part of this work was carried out.

\bibliographystyle{JHEP}
\bibliography{matrix}
\end{document}